\documentclass{jfm}

\usepackage{amsmath,amsfonts,amssymb,stmaryrd,amsbsy,bm}
\usepackage{upmath}
\usepackage{mathtools}
\usepackage{graphicx}
\usepackage{hyperref}
\usepackage{breakurl}
\usepackage{accents}
\usepackage{natbib}
\usepackage{enumitem}

\newcommand{\uu}{{ \bm u}}
\newcommand{\rr}{{ \bm r}}
\newcommand{\pressure}{{p}}
\newcommand{\surface}{S}
\newcommand{\smp}{\varepsilon}
\newcommand{\Y}{{Y_{\ell m}}}
\newcommand{\Yone}{{{\bm Y}^{[1]}_{\ell m}}}
\newcommand{\Ytwo}{{{\bm Y}^{[2]}_{\ell m}}}
\newcommand{\Ythree}{{{\bm Y}^{[3]}_{\ell m}}}
\newcommand{\evec}[1]{\hat{\bm{e}}_{#1}}
\newcommand{\bA}{\bm{\mathcal{A}}}
\newcommand{\A}{{\mathcal{A}}}
\newcommand{\bU}{\bm{\mathcal{U}}}
\newcommand{\U}{{\mathcal{U}}}
\newcommand{\UU}{{ \bm U}}
\newcommand{\cnt}{{\mathcal E}}
\newcommand{\bigO}{{\cal O}}
\newcommand{\inpr}[2]{\left\langle {#1} \middle| {#2} \right\rangle}
\newcommand{\bfY}{{\bm Y}}
\newcommand{\vfY}[2]{{\bfY}^{[#1]}_{#2}}
\newcommand{\vfD}[2]{{\bm D}^{[#1]}_{#2}}
\newcommand{\vfU}[2]{{\bm u}^{[#1]}_{#2}}
\newcommand{\re}[1]{\mathrm{Re}\left\{ #1 \right\} }

\begin{document}

\title{Extrapolation theory for Stokes flow past a deformed sphere}
\author{
Amir Nourhani\aff{1,2,3}
\corresp{\email{Amir.Nourhani@gmail.com}}
Paul E. Lammert\aff{1,2}
\corresp{\email{lammert@psu.edu}},
}

\affiliation{
\aff{1}Center for Nanoscale Science, The Pennsylvania State University, University Park, PA 16802
\aff{2}Department of Physics, Pennsylvania State University, University Park, PA 16802
\aff{3}Department of Physics and Astronomy, Northern Arizona University, Flagstaff, AZ 86011
}

\maketitle

\begin{abstract}
We formulate a method for computing Stokes flow past a highly deformed sphere with arbitrarily defined surface velocity. The fundamental ingredient is an explicit {\it extrapolation operator} extending a velocity field from the surface of a sphere, which is expressed in terms of a complete set of basis Stokes fields for the pressure and velocity derived from scalar and vector spherical harmonics. We present a matrix algebra packaging suitable for numerical computation to arbitrary order in the deformation amplitude (deviation from sphericity). The hydrodynamic force and torque on a deformed sphere with arbitrary surface velocity are expressed in terms of basis field amplitudes, and for the classic problem of a rotating and translating rigid body, we compute explicitly the first order in deformation corrections to the flow field as well as the hydrodynamic force and torque.
\end{abstract}

\section{Introduction\label{sec:intro}}
In his seminal work, Brenner provided a method to solve for the velocity field around a slightly radially deformed sphere with a prescribed surface velocity field~[\cite{Brenner-CES-1964}]. Over the course of half a century, the method has successfully been used to address a broad range of problems, including electrophoresis, diffusiophoresis, osmophoresis, thermophoresis, squirming motion of a sphere, and shear deformation of drops (Appendix \ref{apdx:A} lists papers citing~[\cite{Brenner-CES-1964}], organized by topic). However, the algebraic complication of his method effectively limits its applicability to first order in deformation (i.e., deviation from sphericity). Here, we present a powerful methodology to straightforwardly calculate, essentially by matrix algebra, Stokes flow past a deformed sphere to arbitrary order in the deformation. It is therefore suitable for highly radially-deformed spheres.

Brenner deploys Lamb's general solution~[\cite{lamb-1932}] for the Stokes velocity field,
\begin{equation} 
\uu = \!\!\! \sum_{n=-\infty}^{\infty} \! \left[{\bm \nabla}\! \times \!(\rr \,\chi_n) +  {\bm \nabla}\Phi_n + \frac{(n+3)}{2(n\!+\!1)(2n\!+\!3)}\frac{r^2}{\mu}{\bm \nabla} p_n - \frac{n}{(n\!+\!1)(2n\!+\!3)}\frac{p_n}{\mu}\rr  \right],
 \label{eq:lambvel}
\end{equation}  
where $\rr$ is the position vector. The functions $p_n, \Phi_n,$ and $\chi_n$ are solid spherical harmonics of order $n$ of the form: $H_n(r,\theta,\varphi) \!=\! r^n  \!\sum_{m=0}^{n} P_n^m(\cos\theta)[h_{mn}\cos m\varphi \! + \!\tilde{h}_{mn} \sin m\varphi]$ where $P_n^m$ are the associated Legendre polynomials of the first kind. For a spherical particle of radius $r_0 \!= \!|\rr_0|$, the surface boundary condition for the velocity field $\uu(\rr_0) \!=\! \uu(r_0,\theta,\varphi)$ is used to obtain the expansions $({\rr_0}/{r_0})\cdot  \uu(r_0,\theta,\varphi) \!=\! \sum_{n=1}^\infty \!X_n(\theta,\varphi)$, $-r_0 \bnabla\!\cdot \! \uu(r_0,\theta,\varphi)\!= \!\sum_{n=1}^\infty \!Y_n(\theta,\varphi)$ and $\rr_0\cdot \!\bnabla \times  \uu(r_0,\theta,\varphi) \!= \!\sum_{n=1}^\infty \!Z_n(\theta,\varphi)$ where $X_n, Y_n,$ and $Z_n$ are surface spherical harmonics of the form $H_n(r_0,\theta,\varphi)$. Plugging these expressions into Lamb's solution~(\ref{eq:lambvel}) and imposing the boundary conditions, Brenner obtains the non-zero terms ($n\geq 0$) for a spherical particle, $p_{_{-(n+1)}}\!\! =\! \frac{(2n-1)\mu}{(n+1)r_0} \! \left(\frac{r_0}{r}\right)^{n+1}\!\left[(n\!+\!2)X_n\!+\!Y_n\right]$, $\chi_{_{-(n+1)}} \!\! = \! \frac{1}{n(n+1)} \!\left(\frac{r_0}{r}\right)^{n+1} \!Z_n$, and $\Phi_{_{-(n+1)}} \!\! = \! \frac{r_0}{2(n+1)} \! \left(\frac{r_0}{r}\right)^{n+1}\! \left[n\,X_n\!+\!Y_n\right]$. This treatment for a solid spherical particle is already algebraically complicated and requires applying three operators on the surface velocity field. The calculations for a deformed sphere, as an expansion in powers of the deformation amplitude, are so complicated~[\cite{Brenner-CES-1964}] that it is impractical to proceed even beyond first order. A simpler and more easily implemented method is needed for highly deformed spherical particles.

In this paper, we provide an extrapolation technique to calculate the flow field past a radially deformed sphere, with a specified velocity field on the surface, up to arbitrary perturbation order in deformation amplitude by simple matrix multiplications. In the case of arbitrary geometries which are not amenable to direct analytical calculations with a closed form expression, our method can be used as a semi-analytical technique and has the potential to obtain the numerical elements of the geometry-dependent matrix rapidly by parallel computations.  

In what follows, \S \ref{sec:formulation} formulates the problem and maps the velocity field on a deformed sphere surface $S$ to the surface velocity of a reference sphere $S_0$ (see Fig.~\ref{fig:defsphere}) as an expansion in powers of the deformation amplitude. In \S 3 we introduce the fundamental set of basis Stokes fields and the {\it extrapolation operator}, which directly extrapolates the velocity field on a sphere to the exterior. The velocity field mapping between $S$ and $S_0$, extrapolation operator, and the set of basis Stokes fields are combined in \S 4, to obtain a matrix-based formulation for calculation of fields exterior to a deformed sphere. This formalism is amenable to straightforward numerical implementation to arbitrary order in powers of the deformation amplitude. In \S 5, we obtain the force and torque on a sphere with arbitrary surface velocity field. In \S 6, this is combined with the mapping of \S 2 to obtain the force and torque on a rigidly translating and rotating body to first order in the deformation. Finally, \S 7 concludes with  a discussion of potential applications.

\section{Formulation of the problem}\label{sec:formulation}
\begin{figure}
\begin{center}
\includegraphics[width=0.28\textwidth]{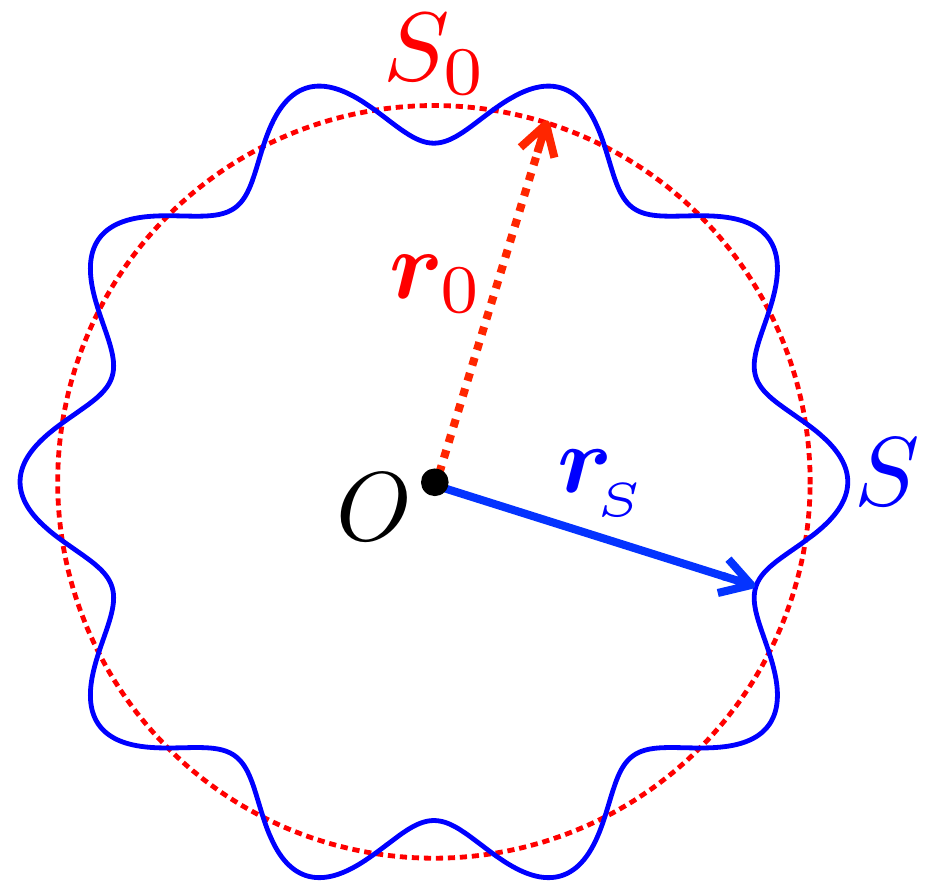}
\end{center}
\caption{
The particle surface $S$ is a radial deformation of the reference sphere $S_0$. 
}
\label{fig:defsphere} 
\end{figure}

The problem is to find velocity ($\uu$) and pressure ($\pressure$) fields of an incompressible Newtonian fluid with viscosity $\mu$ for low Reynolds number flow exterior to a radially deformed spherical surface $S$. We solve the Stokes and continuity equations
\begin{equation}
  \label{eq:Stokes}
\mu \nabla^2 {\uu}  = \bnabla  \pressure\, , \qquad \bnabla \cdot \uu = 0,
\end{equation}
given the surface velocity $\uu(\rr_{_{\!\surface}})$ and far-field values of $\uu$ and $\pressure$. As shown in Fig.~\ref{fig:defsphere}, the surface $S$ is a  radial deformation of a reference sphere $S_0$ of radius $r_0$, specified by
\begin{equation}
\rr_{_{\!\surface}}(\theta,\varphi) = \left[1+\smp \xi(\theta, \varphi) \right]\rr_0(\theta,\varphi), 
\label{eq:defsurf}
\end{equation}
where $\xi$ is an order-one function that describes the particular deviation from sphericity. The perturbative expansion parameter $\smp$ determines 
the deformation amplitude.

The surface velocity field $\uu(\rr_{_{\!\surface}})$ is essentially arbitrary; it could be a simple rigid body motion if $S$ is the bounding surface of a solid particle, but might well be a phoretic slip velocity~[\cite{Nourhani+Lammert-16a}], or a velocity driven by the Marangoni effect~[\cite{young_goldstein_block_1959}] at the surface of a liquid droplet. In our study, we are only concerned with the flow field outside the particle. The velocity on $S$ determines the exterior solution uniquely, assuming the pressure tends to a constant and velocity to zero at infinity (see \S 2.2.1 of [\cite{Kim+Karrila}]). Therefore, by restriction, theoretically we would have a map: (velocity fields on $S$)$\mapsto$(velocity fields on $S_0$). Now, suppose we possessed a method to solve for the flow field exterior to a spherical surface $S_0$, given the velocity on $S_0$. Then, using the preceding map and the solution method we would obtain the velocity field exterior to the deformed sphere.

Pursuing that idea, we first derive a formula for determination of $\uu(\rr_0)$ from $\uu(\rr_{_{\!\surface}})$ and tackle the problem of extrapolation of the Stokes field off of $S_0$ in the next section. To that end, let's write the velocity and pressure fields as expansions in powers of the deformation amplitude $\smp$, 
\begin{equation}
\label{eq:velExpEtraplo}
\uu(\rr) = \sum_{k=0}^{\infty} \smp^k \uu^{(k)} (\rr), \qquad  \pressure(\rr) = \sum_{k=0}^{\infty} \smp^k \pressure^{(k)}(\rr),
\end{equation}
and correspondingly, the velocity field on $S$ as
\begin{equation}
\uu(\rr_{_S}) := \uu_{\!_S} (\theta,\varphi) = \uu_{\!_S}^{(0)} (\theta,\varphi) + \sum_{k = 1}^\infty \smp^k \uu_{\!_S}^{(k)}(\theta,\varphi).
\label{eq:Svelocity}
\end{equation}
Here, we have considered the possibility that the given velocity on $S$ has itself a non-trivial $\smp$-expansion. One such scenario is rigid rotation with angular velocity ${\bm \omega}$ for which the velocity on the surface $S$ is ${\bm \omega}\times \rr_{_S} = {\bm \omega}\times \rr_0 + \smp\, ( \xi {\bm \omega}\times \rr_0)$. Taylor expansion of surface velocity $\uu_{\!_S}$ about the reference sphere $S_0$ yields $\uu_{\!_S}  =  \uu(\rr_0) + \sum_{q=1}^\infty  \frac{1}{q!} \smp^q (r_0\xi)^q \left(\partial^q \uu / \partial r^q \right)|_{r=r_0}$. Combining this Taylor expansion and the perturbation expansion (\ref{eq:Svelocity}), we obtain the velocity expansion on the reference sphere $S_0$ as
\begin{align}
\uu^{(0)}(\rr_0)  =\uu_{\!_S}^{(0)},  \qquad \uu^{(i)}(\rr_0)  =\uu_{\!_S}^{(k)} -  \sum_{q=1}^k \frac{1}{q!}(r_0\xi)^q \left( \frac{\partial^q \uu^{(k-q)}}{\partial r^q} \right)\Big|_{r=r_0}\ \text{for}\ k\geq 1.
\label{eq:generalBC}
\end{align}
As usual, the question of convergence of the $\smp$-expansion is completely unclear~[\cite{Brenner-CES-1964}], but also not really the right question, which is whether a few terms give a physically legitimate answer.

\section{Extrapolation Operator for Spherical Particle\label{sec:extrapolationoperator}}

A central component of our formalism is an ``extrapolation operator,'' $${\cal E}: \text{velocity field on a reference sphere} \mapsto \text{velocity field off the sphere surface},$$ that extrapolates an arbitrary velocity field on a sphere to all of space and provides a solution to the Stokes equations when paired with an appropriate pressure field.

A Stokes pressure field, satisfying the Laplace equation $\nabla^2 \pressure = 0$, and bounded at infinity can be expanded in the basic solutions $r^{-(\ell+1)} \Y$ where $\Y$ is a scalar spherical harmonic and $r$ is the radial coordinate in the spherical coordinate system. Furthermore, the expansion coefficients can be extracted from the boundary condition on the sphere by virtue of the orthogonality of the scalar spherical harmonics. Analogously, to obtain the Stokes velocity field we need the {\em vector spherical harmonics}
\begin{align}
\Yone = \Y \evec{r}, \quad 
\Ytwo = r \bnabla \Y, \quad
\Ythree = \rr \times\bnabla \Y,
\label{eq:defVSH}
\end{align} 
as a complete basis for vector functions on a sphere. Their orthogonality relation is,
\begin{equation}
 \inpr{ \vfY{\alpha}{\ell m} }{ \vfY{\beta}{\ell^\prime m^\prime} }
= 
\int_S {\bm Y}^{[\alpha]*}_{\ell m} \cdot {\bm Y}^{[\beta]}_{\ell^\prime m^\prime}  \, d\Omega
=
[ \ell (\ell+1) (1 - \delta_{1\alpha}) +   \delta_{1\alpha}] \delta_{\alpha\beta} \delta_{\ell \ell^\prime} \delta_{m m^\prime},
\label{eq:HilbertOrthog}
\end{equation}
where $\delta_{ij}$ is the Kronecker delta, in $\inpr{ {\bm V} }{ {\bm W} } = \int_S  {\bm V}^* \cdot {\bm W} d\Omega$, the integration domain is the surface of a sphere, and the vector ${\bm V}^*$ is the complex conjugate of ${\bm V}$. 

A Stokes velocity field could be expanded as $\sum_{\alpha=1}^3 g_{\ell, m}^{[\alpha]} (r) \, {\bm Y}^{[\alpha]}_{\ell m}$~[\cite{Olla-JPAMG-1997, Olla-MFm-2011}], but individual ${\bm Y}_{\ell m}^{[\alpha]}$ by themselves are a suboptimal basis for our purpose since the radial function coefficients $g_{\ell m}^{[\alpha]}$ will not be homogeneous functions in $r$. Enforcing the simple powers $g^{[\alpha]}_{\ell, m}(r) \propto r^{-\beta}$ we obtain a fundamental basis set for Stokes velocity and pressure fields as,
\begin{subequations}
\begin{align}
& \uu_{\ell m}^{[1]} (\rr) = \left[\frac{r_0}{r}\right]^{\ell} \left[ \ell (\ell+1) \Yone - (\ell -2) \Ytwo \right],
& &
\pressure_{\ell m}^{[1]}(\rr)  = 2\left[\frac{\mu}{r_0}\right] \left[\frac{r_0}{r}\right]^{(\ell+1)}\ell ( 2 \ell -1)   \, \Y,
\label{eq:velterm1}
\\
& \uu_{\ell m}^{[2]}  (\rr)
= \left[\frac{r_0}{r}\right]^{(\ell+2)} \left[ - (\ell+1) \Yone + \Ytwo \right],
& & 
\pressure_{\ell m}^{[2]}  (\rr) = 0,
\label{eq:velterm2}
\\
& \uu_{\ell m}^{[3]}  (\rr) 
= \left[\frac{r_0}{r}\right]^{(\ell+1)} \Ythree \
 = \frac{\bm r}{r_0} \times \uu_{\ell m}^{[2]}  (\rr),
& &
\pressure_{\ell m}^{[3]}  (\rr) = 0.
\label{eq:velterm3}
\end{align}
\label{eq:velterms}
\end{subequations}
We refer to these as type-1, 2 or 3 Stokes basis fields. An arbitrary Stokes velocity field is then expanded in terms of the Stokes basis as $\uu = \sum_{\alpha=1}^3\sum_{\ell, m} C_{\ell m}^{[\alpha]} \uu_{\ell m}^{[\alpha]}$, with the corresponding pressure field $\pressure = \sum_{\ell, m} C_{\ell m}^{[1]} \pressure_{\ell m}^{[1]}$.

As the Stokes basis fields (\ref{eq:velterm1}) -- (\ref{eq:velterm3}) are not orthogonal over a sphere, the expansion coefficients $C_{\ell m}^{[\alpha]} =  \langle {\bm D}_{\ell m}^{[\alpha]} | \uu(\rr_0)  \rangle$ are extracted by use of dual vectors ${\bm D}_{\ell m}^{[\alpha]}$ satisfying
\begin{equation}
\inpr{ \vfD{\alpha}{\ell m}}{ \vfU{\beta}{\ell^\prime m^\prime}(\rr_0) }= \delta_{\alpha \beta} \delta_{\ell \ell^\prime} \delta_{m m^\prime}.
\label{eq:orthoDu}
\end{equation}
Using the above orthogonality relation and the Stokes basis fields (\ref{eq:velterms})  we obtain 
\begin{align}
& {\bm D}_{\ell m}^{[1]}  = {1 \over 2\ell(\ell+1)} \left[\ell\, \Yone +  \Ytwo \right], \quad
{\bm D}_{\ell m}^{[2]}   = {1 \over 2(\ell+1)} \left[(\ell-2)\, \Yone +  \Ytwo \right], \quad
\nonumber \\
& {\bm D}_{\ell m}^{[3]}  = {1 \over \ell(\ell+1)} \Ythree = 2 \, \evec{r} \times {\bm D}_{\ell m}^{[1]} = 2 \ell^{-1}\, \evec{r} \times {\bm D}_{\ell m}^{[2]}.
\end{align}
Combining the Stokes basis with the dual vectors, the extrapolation of a velocity field $\uu(\rr_0)$ on the sphere of radius $r_0$ is 
\begin{align}
\uu(\rr) & = 
{\cal E}\left[\uu(\rr_0)\right] = \sum_{\alpha=1}^3\sum_{\ell, m} \inpr{ \vfD{\alpha}{\ell m} }{ \uu(\rr_0) } \vfU{\alpha}{\ell m}(\rr),
\label{eq:velocityexpansion}
\\
\pressure(\rr) & = {\cal P}\left[\uu(\rr_0)\right] = \sum_{\ell, m} \inpr{ \vfD{1}{\ell m} }{ \uu(\rr_0) } p_{\ell m}^{[1]} (\rr),
\label{eq:pressureexpansion}
\end{align}
which defines the extrapolation operators ${\cal E}$ and ${\cal P}$ for velocity and pressure fields, respectively. Here, we obtained the expansion coefficients directly from the value of the velocity field of the sphere surface, compared to Brenner's method~[\cite{Brenner-CES-1964}], which required first applying three operators to the surface velocity field. Since the basis fields are proportional to negative powers of $r$, if the expansions converge on $S_0$, they will converge outside. Convergence inside the sphere is more delicate. However, with only a finite number of terms, there will only be a singularity at the origin.

\section{Deformed Sphere and Recursive Matrix Formalism}

We are now ready to exploit the mapping (\ref{eq:generalBC}) and the extrapolation operator (\ref{eq:velocityexpansion}) to calculate the flow field around the radially deformed sphere surface specified by expression (\ref{eq:defsurf}). The calculations are cast in a convenient matrix form by organizing the allowed $(\ell,m)$ values into a linear order indexed according to
\begin{equation}
J = \ell(\ell+1)+m+1,
\quad
\text{and inversely}
\quad
\ell = \lfloor \sqrt{J-1}\rfloor, \;\; m = J - \ell(\ell+1)-1, 
\label{eq:indexmapping}
\end{equation}
where $\lfloor x\rfloor$ is the largest integer not exceeding $x$.
The relationship (\ref{eq:indexmapping}) between
$J$, $\ell$ and $m$ (or $J^\prime$, $\ell^\prime$ and $m^\prime$)
is implicit whenever they appear together.
Then, the basis fields can be collected into three vectors ($\alpha = 1,2,3$) as (superscript $T$ stands for transpose) 
\begin{equation}
\bU^{[\alpha]\,T} \!\! = \! \left[\U^{[\alpha]}_1,\, \U^{[\alpha]}_2, \, \U^{[\alpha]}_3, \, \cdots\right] \! = \!
\left[\uu_{0,0}^{[\alpha]},\, \uu_{1,-1}^{[\alpha]}, \, \uu_{1,0}^{[\alpha]}, \, \uu_{1,1}^{[\alpha]},\, \cdots \uu_{\ell,-\ell}^{[\alpha]}, \cdots, \uu_{\ell,0}^{[\alpha]}, \cdots, \uu_{\ell,\ell}^{[\alpha]}  \cdots \right].
\label{eq:bigUvector}
\end{equation}

For each perturbation order $\smp^k$, we define the coefficient vectors $\A^{[\alpha]}$ and $\A_{_S}^{[\alpha]}$ consisting of the expansion coefficients of velocity fields on the reference sphere $S_0$ and deformed sphere $S$, respectively, as
\begin{equation}
(\A^{[\alpha](k)})_J = \inpr{ \vfD{\alpha}{\ell m} }{ \uu^{(k)}(\rr_0) },
\qquad
(\A_{_S}^{[\alpha](k)})_J = \inpr{ \vfD{\alpha}{\ell m} }{ \uu_{_S}^{(k)} }.
\label{eq:A matrices}
\end{equation}
The $\bigO(\smp^k)$ velocity field is then
\begin{equation}
\uu^{(k)}(\rr) = {\cal E}[\uu(\rr_0)] = \sum_{\alpha=1}^3\sum_{J}  \inpr{ \vfD{\alpha}{\ell m} }{ \uu^{(k)}(\rr_0) }\uu_{J}^{[\alpha]} (\rr) = \sum_{\alpha=1}^3 \left(\bA^{[\alpha](k)}\right)^T  \bU^{[\alpha]}, 
\end{equation}
and the complete velocity field,
\begin{equation}
\uu (\rr) = \sum_{k\ge 0} \smp^k  \sum_{\alpha = 1}^3 \left( \bA^{[\alpha](k)} \right)^T \bU^{[\alpha]}.
\label{eq:velocity field matrix form}
\end{equation}

Then, to obtain the velocity field $\uu (\rr)$, we need to obtain  the coefficient vectors $\bA^{[\alpha](k)}$ using the definition (\ref{eq:A matrices}) and the  recursive formulas (\ref{eq:generalBC}). For this purpose, it is convenient to define the notation $n^{[\alpha]}(J)$ for the exponents in $\uu_{\ell,m}^{[\alpha]}(\rr) \sim r^{-n^{[\alpha]}(J)}$ as 
\begin{equation}
n^{[1]}(J) = \ell,\quad n^{[2]}(J) = \ell+2, \quad n^{[3]}(J) = \ell+1.
\end{equation}
With that, we define square matrices ${\bm \Xi}^{[\beta,\alpha]\{q\}}$ having components
\begin{equation}
{\bm \Xi}^{[\beta,\alpha]\{q\}}_{J,J^\prime} = (-1)^q \frac{(n^{[\beta]}(J)+q-1)!}{q! (n^{[\beta]}(J)-1)!} \, \inpr{ {\bm D}_{J^\prime}^{[\alpha]} }{ \xi^q \,\uu_{J}^{[\beta]} (\rr_0) }.
\label{eq:Xi matrices}
\end{equation}
The recursion relations~(\ref{eq:generalBC}) for mapping velocity field on the deformed sphere $S$ to the velocity field on the reference sphere $S_0$ are now written in the recursive matrix form as
\begin{equation}
\bA^{[\alpha](k)} = \bA_{_S}^{[\alpha](k)}  - \sum_{q=1}^k \sum_{\beta =1}^3 \left(\bA^{[\beta](k-q)}\right)^T {\bm \Xi}^{[\beta,\alpha]\{q\}}
\label{eq:matrix recursion formulas}
\end{equation}
The matrices ${\bm \Xi}^{[\beta,\alpha]\{q\}}$ depend on the {\em geometry} of the problem. Only the vectors $\bA_{_S}^{[\alpha](k)}$  directly depend on  the velocity boundary condition on the deformed surface $S$.

Equations (\ref{eq:bigUvector}), (\ref{eq:A matrices}), (\ref{eq:velocity field matrix form}), (\ref{eq:Xi matrices}) and (\ref{eq:matrix recursion formulas}) give the complete solution. Restricting the index $J$ to run over a finite set,  the formalism is amenable to numerical execution.

\section{Force and torque on the particle}

The homogeneous radial dependence of the terms in expansions (\ref{eq:velocityexpansion}) and (\ref{eq:pressureexpansion}) is advantageous when considering the net hydrodynamic force and torque on the particle. Since that force is equal to the momentum flux through any closed surface surrounding the particle, it can be calculated using an infinitely large sphere, in which case only the component of velocity decaying as $r^{-1}$ or pressure decaying as $r^{-2}$ can possibly contribute to the force calculation. But only the type-1 $\ell=1$ Stokes basis fields have such property, hence the  force can be computed from just the three velocity terms $\sum_{m=-1}^1\langle  \vfD{1}{1 m} | \uu(\rr_0) \rangle \uu_{1,m}^{[1]}.$ Similarly, a finite nonzero torque can only arise from a velocity (pressure) component decaying as $r^{-2}$ ($r^{-3}$). Together with the observation that torque, being a vector, transforms under rotations as $\ell=1$, this shows that the torque is determined by the three velocity terms $\sum_{m=-1}^1\langle \vfD{3}{1 m}  | \uu(\rr_0) \rangle \uu_{1,m}^{[3]}$. Now note that the constant vectors $\vfD{1}{1 m}$ are a basis for uniform vector fields over the sphere:
\begin{equation}
{\bm D}_{1, 0}^{[1]} = \frac{1}{8} \sqrt{\frac{3}{\pi}}\evec{z}, \qquad {\bm D}_{1, \pm 1}^{[1]} = \mp \frac{1}{8} \sqrt{\frac{3}{\pi}}\evec{\pm}
\quad \text{with} \quad \evec{\pm} = \frac{1}{\sqrt{2}}(\evec{x} \pm i \evec{y}),
\label{eq:D^1_1}
\end{equation}
where $(\evec{x},\evec{y},\evec{z})$ are unit vectors of a Cartesian coordinate system. Therefore, the components $\langle \vfD{1}{1 m} | \uu(\rr_0) \rangle$ extract the the average of ${\bm u}({\bm r}_0)$ over the reference sphere and the net force ${\bm F}$ exerted by the fluid on the particle is proportional to the average of $\uu(\rr_0)$ over the sphere surface, that is, $\overline{\uu(\rr_0)} := (4\pi)^{-1}\int_S \uu(\rr_0)\, d\Omega$, as
\begin{equation}
{\bm F} = -6\pi\mu \,r_0\, \overline{\uu(\rr_0)}
\label{eq:net force}
\end{equation}
As for the torque, note that since $\vfD{3}{1 m} = 2\evec{r}\times \vfD{1}{1 m}$, for the coefficients of the contributing velocity terms we have $\langle \vfD{3}{1 m} | \uu(\rr_0)\rangle = - 2 \langle \vfD{1}{1 m} |  \evec{r}   \times \uu(\rr_0) \rangle$, and thus,  the torque ${\bm L}$ is proportional to the average of $\evec{r}   \times {\bm u}({\bm r}_0)$ over the reference sphere, as
\begin{equation}
{\bm L} = - 12\pi\mu \, r^2_0  \,\,\overline{ \evec{r}   \times \uu(\rr_0)}.
\label{eq:net torque}
\end{equation}
For a rigid body rotation of an sphere with angular velocity ${\bm \omega}$, the surface velocity is $\uu(\rr_0) = {\bm \omega} \times \rr_0$, yielding $\overline{ \evec{r}   \times \uu(\rr_0)} = \frac{2}{3}r_0{\bm \omega}$, and we obtain the well-know expression  ${\bm L} = - 8\pi\mu\,r_0^3\,{\bm \omega}$ for the torque exerted by fluid on the rotating rigid sphere.

To connect with the matrix formalism of \S 4, using the expression (\ref{eq:velocity field matrix form}), the identities
\begin{align}
&\overline{\uu^{[\alpha]}_{\ell m}(\rr_0)} =2 \sqrt{\frac{1}{3\pi}} \delta_{\alpha, 1} \delta_{\ell, 1}\left[- \evec{+}  \delta_{m, +1}+\evec{-}  \delta_{m, -1}+\evec{z}  \delta_{m, 0}\right],
\label{eq:average velocity}
\\
&\overline{\evec{r}\times \uu^{[\alpha]}_{\ell m}(\rr_0)}= - \sqrt{\frac{1}{3\pi}} \delta_{\alpha, 3} \delta_{\ell, 1}\left[- \evec{+}  \delta_{m, +1}+\evec{-}  \delta_{m, -1}+\evec{z}  \delta_{m, 0}\right],
\label{eq:average r x velocity}
\end{align}
and the $J$ indices 2, 3 and 4 corresponding to $(\ell, m)$ pairs $(1,-1)$, $(1,0)$ and $(1,+1)$, respectively, 
according to relation (\ref{eq:indexmapping}), we obtain 
\begin{align}
{\bm F} 
&=
-4\sqrt{3\pi}\mu \,r_0\, \sum_{k\ge 0} \varepsilon^k  \left[ \left( \bm{\mathcal{A}}^{[1](k)} \right)_2 \evec{-} + \left( \bm{\mathcal{A}}^{[1](k)} \right)_3 \evec{z}-  \left( \bm{\mathcal{A}}^{[1](k)} \right)_4\evec{+} \right],
 \\ 
{\bm L} &= 4\sqrt{3\pi}\mu \,r_0^2\,  \sum_{k\ge 0} \varepsilon^k   \left[ \left( \bm{\mathcal{A}}^{[3](k)} \right)_2 \evec{-}+ \left( \bm{\mathcal{A}}^{[3](k)} \right)_3\evec{z} - \left( \bm{\mathcal{A}}^{[3](k)} \right)_4 \evec{+}  \right]. 
\end{align}
For each order of perturbation, $k$, we only need three elements of the vectors $\bm{\mathcal{A}}^{[1](k)}$ and $\bm{\mathcal{A}}^{[3](k)}$ to calculate force and torque, respectively.

\section{Rigid-body motion of a deformed sphere}

\subsection{Definition of the problem}

Here we treat the classic problem of rigid body motion at the surface of the radially deformed sphere, $\rr_{_{\!\surface}}(\theta,\varphi) = \left[1+\smp \xi(\theta, \varphi) \right]\rr_0(\theta,\varphi)$ as in (\ref{eq:defsurf}). The surface velocity is
\begin{equation}
\uu(\rr_{_S}) =  \UU + {\bm \omega}\times \rr_{_S} =  \UU +  {\bm \omega}\times \rr_0 + \smp\,\xi {\bm \omega}\times \rr_0
\label{eq:rigid body motion}
\end{equation}
where $\UU$ and ${\bm \omega}$ are the constant translational and angular velocities. Given the form (\ref{eq:D^1_1}) of dual vectors $\vfD{1}{1 m}$,
it is convenient to expand the translational velocity as
\begin{equation}
\UU = U_+ \evec{-} + U_- \evec{+} + U_z \evec{z}
\quad
\text{with}
\quad
U_{\pm} = U_{\mp}^*  = \frac{1}{\sqrt{2}}(U_{x} \pm i U_{y}) = \UU\cdot\evec{\pm}.
\label{eq:velocityComplexCoord}
\end{equation}
We use a similar expansion for the angular velocity ${\bm \omega}$. 
In what follows we calculate to ${\cal O}(\smp)$ the translational flow field $\uu_\text{trs}$ arising from $\UU$ and the rotational flow field $\uu_\text{rot}$ arising from ${\bm \omega}$, separately. Exploiting the linearity of the Stokes and continuity equations, the case of general rigid body motion is then obtained simply by superposition.

\subsection{Translational motion of a radially deformed sphere particle up to $\bigO(\smp)$}

To zero-th order in $\smp$, the velocity on the reference sphere is simply $\uu_{\text{trs}}^{(0)}(\rr_0) = \UU$. According to (\ref{eq:D^1_1}), this can be written in terms of the constant basis $\vfD{1}{1 m}$, $m=-1,0,1$.
The normalizations in (\ref{eq:HilbertOrthog}) yield
$
\langle\vfD{1}{1 m}|\vfD{1}{1 m^\prime}\rangle  \!= \!\langle\vfD{2}{1 m}|\vfD{2}{1 m^\prime}\rangle \! = \! 3 \langle\vfD{1}{1 m}|\vfD{2}{1 m^\prime}\rangle  \! = \! \frac{3}{16}\delta_{m,m^\prime}
$,
and all type-1 and type-2 fields are orthogonal to type-3 fields. Thus, the general extrapolation formula (\ref{eq:velocityexpansion}) yields
\begin{align}
\uu_\text{trs}^{(0)}(\rr) = \sqrt{3\pi \over 4} \,\left[2 \text{Re} \left\{U_+ \left(\uu_{1,-1}^{[1]} + \frac{1}{3}\uu_{1,-1}^{[2]}\right) \right\}+  U_z \left(\uu_{1,0}^{[1]} + \frac{1}{3}\uu_{1,0}^{[2]}\right)
\right], 
\label{eq:constlinvel}
\end{align}
where we have used the real part operator $\text{Re}$ and the complex conjugation formula ${\bm Y}^{[\alpha]}_{\ell,-m} = (-1)^m {\bm Y}^{[\alpha]*}_{\ell,m} $.

Following Eq.~(\ref{eq:generalBC}), with $- r_0\partial_r[ \uu_{1,m}^{[1]}(\rr) + \frac{1}{3} \uu_{1,m}^{[2]}(\rr)]_{r=r_0}= 2 {\bm Y}_{1, m}^{[2]}$, the velocity on the reference sphere is
\begin{align}
\uu_{\text{trs}}(\rr_0) &= \UU + \smp \sqrt{3\pi} \,\xi \left( 2 \text{Re}\{U_+ {\bm Y}_{1,-1}^{[2]} \}  + U_z {\bm Y}_{1,0}^{[2]}\right) + \bigO(\smp^2)
\nonumber \\
&
= \UU + \smp \frac{3}{2} \,\xi \left[ \UU - \evec{r}(\evec{r}\cdot\UU) \right]+ \bigO(\smp^2).
\end{align}
Straightforward application of the general formula (\ref{eq:velocityexpansion}) to $\uu_\text{trs}^{(1)}(\rr) $ gives the $\cal{O}(\smp)$ term
\begin{equation}
\uu_{\text{trs}}^{(1)}(\rr) = \sqrt{3\pi} \sum_{\ell, m} \left\{{\cal G}_{\ell, m}^{[2]}(\xi \UU) \left[ \uu_{\ell, m}^{[1]}(\rr) + \ell \uu_{\ell, m}^{[2]}(\rr)\right] + 2 {\cal G}_{\ell, m}^{[3]}\!(\xi\UU) \, \uu_{\ell, m}^{[3]}(\rr) \right\}
\end{equation}
where for arbitrary real vector field ${\bm W}$ on the sphere,
\begin{align}
{\cal G}_{\ell, m}^{[\beta]}({\bm W})  &= \frac{1}{2\ell(\ell+1)} \left[\inpr{ \vfY{\beta}{\ell,m} }{2\, \re{ W_+  \vfY{2}{1,-1} } }+ \inpr{ \vfY{\beta}{\ell,m} }{ W_z \, \vfY{2}{1,0} }\right]
\nonumber \\
& = \frac{1}{2\ell(\ell+1)} \sqrt{\frac{3}{4\pi}}\inpr{\vfY{\beta}{\ell,m}}{({ \cal I}-\evec{r}\evec{r}) \cdot {\bm W} }
\label{eq:Gs}
\end{align}
where ${\cal I}$ is the identity tensor.
The net force on the particle depends upon the velocity average over a sphere, according to (\ref{eq:net force}). Using (\ref{eq:average velocity}) and the precise expressions for the $\vfY{2}{1,m}$'s, after some manipulations we obtain
\begin{equation}
{\bm F} = -6\pi\mu r_0 \left[ {\bm U} + \frac{3}{8\pi} \smp\int \xi ({\cal I}-\evec{r}\evec{r})\cdot\UU \, d\Omega\right] +\bigO(\smp^2) 
\end{equation}

 \subsection{Rotational motion of a deformed spherical particle up to $\bigO(\smp)$}

For rigid body rotation with angular velocity ${\bm \omega}$, the velocity at the surface $S$ is given by Eq. (\ref{eq:rigid body motion}) with $\UU=0$. Use of Eq.~(\ref{eq:generalBC}) gives something a velocity on the reference sphere with an extra term compared to the translational case:
\begin{align}
\uu^{(0)}_\text{rot}(\rr_0) &= {\bm \omega}\times \rr_0, 
\qquad
\uu^{(1)}_\text{rot}(\rr_0) = \xi \left[ {\bm \omega}\times \rr_0 - r_0 \frac{\partial}{\partial r}\uu_\text{rot}^{(0)}\Big|_{r=r_0}\right].
\label{eq:rotBC}
\end{align}

To extrapolate the $\bigO(\smp^0)$ field, we use the identity $\inpr{{\bm V}\times {\bm r}_0}{{\bm W}} = \inpr{{\bm V}}{{\bm r}_0 \times{\bm W}}$,  along with $\rr_0 \times \Yone = 0$,  $ \rr_0 \times \Ytwo = r_0 \Ythree$, and $\rr_0 \times \Ythree = -r_0 \Ytwo$.
Applying these with straightforward algebra to the general extrapolation formula (\ref{eq:velocityexpansion}) results in
\begin{equation}
\uu^{(0)}_\text{rot}(\rr) = -2 r_0 \sqrt{\pi \over 3} \, \left[ 2 \text{Re} \left\{\omega_+ \uu_{1,-1}^{(3)} \right\} + \omega_z \uu_{1,0}^{(3)}\right]. 
 \label{eq:zeroOrderRigidRot}
\end{equation}
Then, using $- r_0 \partial_r \uu^{(0)}\big|_{r=r_0} =  2 \uu^{(0)}(r_0)$, we obtain
$
\uu_\text{rot}(\rr_0) \!=\! (1 \!+ \!3 \smp \xi) \, {\bm \omega}\times \rr_0 \!+\!\bigO(\smp^2).
$
Manipulations similar to those used for the $\bigO(\smp^0)$ rotational term and the $\bigO(\smp^1)$ translational term yield the expression
\begin{align}
 \uu^{(1)}_\text{rot}(\rr) & = 6 r_0 \sqrt{\pi \over 3} \, \sum_{\ell, m} \left({\cal G}_{\ell, m}^{[3]}(\xi {\bm \omega}) \left[ \uu_{\ell, m}^{[1]}(\rr) + \ell \uu_{\ell, m}^{[2]}(\rr)\right] - 2 {\cal G}_{\ell, m}^{[2]}(\xi {\bm \omega}) \uu_{\ell, m}^{[3]}(\rr) \right),
\label{eq:1st order velocity from rotation}
\end{align}
for $\uu^{(1)}_\text{rot}(\rr_0) = 3 \cnt [\xi  {\bm \omega}\times \rr_0]$. The auxiliary functions ${\cal G}_{\ell, m}^{[\alpha]}$ are defined in Eq. (\ref{eq:Gs}).

Now using (\ref{eq:net torque}) and (\ref{eq:average r x velocity}) in conjunction with (\ref{eq:1st order velocity from rotation}), and some manipulation yields
\begin{equation}
{\bm L} = - 8\pi\mu \, r_0^3 \left[{\bm \omega} + \frac{9}{8\pi} \smp \int \xi ({\cal I}-\evec{r}\evec{r}) \cdot {\bm\omega} \, d\Omega \right] +\bigO(\smp^2).
\end{equation}

\section{Conclusion}

Methods based on vector spherical harmonics and the notion of an extrapolation operator appear to be well suited to perturbative expansions of exterior Stokes flow problems. However, we emphasize this approach is certainly not limited to pencil-and-paper calculations with a spherical boundary. We have shown how deviation from sphericity of the bounding surface can be handled to first order as well as how the calculations can be organized for computer implementation to high perturbative order. Beyond spherical  microswimmers~[\cite{Nourhani+-15c-PRE}], a significant potential use of the formalism is to study of geometrical effects for microswimmers and nanomotors of highly non-spherical shape~[\cite{laboraving2016}].

\section{Acknowledgment}
We acknowledge funding by the National Science Foundation under Grant No. DMR-1420620 through the Penn State Center for Nanoscale Science.

\appendix
\section{Applications of Brenner's formalism
\label{apdx:A}
}

As indicated in the introduction, Brenner's method has been applied to a large variety of problems. The following is a list of all the papers we found citing~[\cite{Brenner-CES-1964}]:

\begin{itemize}[labelsep=6pt,itemsep=4pt]
\item Electrophoresis:~\cite{Anderson-JCIS-1984},~\cite{Keh-JFM-1985},~\cite{Chen-AIChE-1988},
\item[] \cite{Feng-PT-2002}
\item  Diffusiophoresis:~\cite{Keh-PF-1995}
\item Osmophoresis:~\cite{Keh-IJMF-2000}
\item Thermophoresis:~\cite{Mackowski-JCIS-1990},~\cite{Senchenko-FP-2007},~\cite{Chen-AS-2002},
\item[] \cite{Mohan-JAM-2006},~\cite{Chang-JAS-2010}
\item Thermocapillary motion:~\cite{Anderson-IJMF-1985},~\cite{Chen-L-2003}
\item Rheology:~\cite{Schowalter-JCI-1968},~\cite{Brenner-ARFM-1970}
\item Effects of surfactant on dynamics of compound droplets:~\cite{Mandal-JFM-2016}
\item Squirming motion of a sphere:~\cite{Pak-JEM-2014}
\item Two unequal viscous drops in Stokes flow:~\cite{Fuentes-PF-1989}
\item Deformed sphere with slip velocity:~\cite{Jia-ASME-2005},~\cite{Chang-JCIS-2009}
\item Fluid drops in linear and quadratic flow:~\cite{Yang-KJCE-1989}
\item  Nanorheology of viscoelastic shells:~\cite{Kuriabova-PRE-2008}
\item Effective viscosity of suspensions and emulsions:~\cite{Yaron-RA-1972},
\item[] \cite{Chaitey-JCS-1965}
\item Capillary propulsion:~\cite{Pak-Lauga-2014}
\item Shear deformation of drops:~\cite{Chaffey-JCI-1967}
\end{itemize}

\bibliographystyle{jfm}

\end{document}